# Pion, Kaon, and (Anti-) Proton Production in Au+Au Collisions at $\sqrt{s_{NN}}$ = 62.4 GeV

Ming Shao[1,2] for the STAR Collaboration

[1]*University of Science & Technology of China, Anhui 230027, China*
[2]*Brookhaven National Laboratory, Upton, New York 11973, USA*

PACS: 25.75.Dw, 12.38.Mh

**Abstract**. We report on preliminary results of pion, kaon, and (anti-) proton transverse momentum spectra ($-0.5 < y < 0$) in Au+Au collisions at $\sqrt{s_{NN}}$ = 62.4 GeV using the STAR detector at RHIC. The particle identification (PID) is achieved by a combination of the STAR TPC and the new TOF detectors, which allow a PID coverage in transverse momentum ($p_T$) up to 7 GeV/$c$ for pions, 3 GeV/$c$ for kaons, and 5 GeV/$c$ for (anti-) protons.

## 1. Introduction

In 2004, a short run of Au+Au collisions at $\sqrt{s_{NN}}$ = 62.4 GeV was accomplished, allowing to further study the many interesting topics in the field of relativistic heavy-ion physics.

The measurements of the nuclear modification factors $R_{AA}$ and $R_{CP}$ [1][2] at 130 and 200 GeV Au+Au collisions at RHIC have shown strong hadron suppression at high $p_T$ for central collisions, suggesting strong final state interactions (in-medium) [3][4][5]. At 62.4 GeV, the initial system parameters, such as energy and parton density, are quite different. The measurements of $R_{AA}$ and $R_{CP}$ up to intermediate $p_T$ and the azimuthal anisotropy dependence of identified particles at intermediate and high $p_T$ for different system sizes (or densities) may provide further understanding of the in-medium effects and further insight to the strongly interacting dense matter formed in such collisions [6][7][8][9].

A scaling behavior according to the number of constituent quarks in a hadron was found to dominate the hadron production and the elliptic flow at 200 GeV Au+Au collisions [10][11][12], and was successfully explained by the coalescence/recombination model [6][7][8][9]. In the case of the $p/\pi$ ratio, the coalescence model [13] predicts that after a maximum at $p_T \sim$ 2-3 GeV/c, the $p/\pi$ ratio should decrease at larger $p_T$ [8][9]. It is then interesting to check the particle species dependence by measuring the $R_{CP}$ of identified mesons (pion, kaon) and baryons (proton) and the $p/\pi$ prediction at 62.4 GeV Au+Au collisions, a system with significant lower parton density.

In this paper we will present a method to extend the particle identification range at the STAR experiment. Based on this method, the pion, kaon and (anti-) proton spectra at 62.4 GeV Au+Au collisions are presented.

## 2. Experimental Setup (and Data Set)

The detector used for these studies was the Solenoidal Tracker at RHIC (STAR) [14]. The main tracking device is the Time Projection Chamber (TPC). By requiring the tracks of charged particles to have at least 15 out of a maximum of 45 hits in the TPC, a dE/dx resolution of ~ 8% can be achieved. Detailed descriptions of the TPC have been presented in [15]. Such a good resolution allows particle identification at intermediate and high $p_T$, in addition to the conventional particle identification at low $p_T$.

A time-of-flight detector (TOFr) based on multi-gap resistive plate chambers (MRPC) [16] was installed in STAR in 2003. It covers 1/60 in azimuth and $-1 < |\eta| < 0$ in pseudo-rapidity at a radius of ~ 220 cm. Two identical pseudo-vertex position detectors (pVPD) were installed to record the starting time for the TOFr, each 5.4 m away from the TPC center along the beam line. Each pVPD covers ~ 19% of the total solid angle in $4.43 < |\eta| < 4.94$ [17].

The data set used in this paper is from Au+Au collisions at $\sqrt{s_{NN}}$ = 62.4 GeV. There are ~7.2M events that pass the standard STAR minimum bias trigger, from which ~2.2M events have TOFr matched hits. The event vertex is required to be within 30 cm from the center of STAR detector. Other cuts applied to the analysis: a global distance of closest approach (DCA) less than 3cm, a rapidity range of $-0.5 < y < 0$, and a TOFr hit position within readout strip geometry bounds. Some further cuts to the TOFr signals (such as amplitude, timing, and etc.) are also applied to assure good TOFr signal quality.

## 3. Particle Identification Methods

The TOFr has an intrinsic time resolution of ~ 85 ps [18], and for the start counter pVPD, it is ~ 55 ps. Thus, the TOF system alone can identify kaons from pions up to $p_T$ ~ 1.6 GeV/$c$ and (anti-) protons from pions and kaons up to $p_T$ ~ 3.0 GeV/$c$ (see Figure 1).

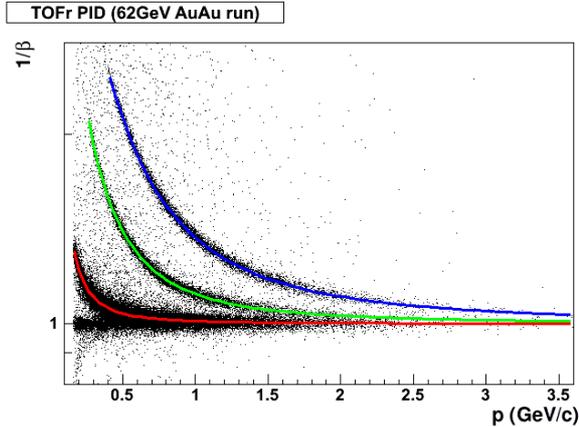

Figure 1: 1/β vs. momentum for pion, kaon and (anti-) proton from 62.4 GeV Au+Au collisions. The separation between pions and kaons is achieved up to $p_T$ ~ 1.6 and the separation between kaons and (anti-) protons and is achieved up to $p_T$ ~ 3.0 GeV/c.

The specific energy loss dE/dx has long been used for low $p_T$ particle identification. However, it is also possible to separate pions, kaons and (anti-) protons in a much higher momentum range when we combine the TOF information with the dE/dx measurement from the TPC. Figure 2(a) shows the hadron distribution as a function of

$n\sigma_\pi$ and $m^2$, where $n\sigma_\pi$ is defined as $n\sigma_\pi = \ln(A/B)/\sigma_{TPC}(L)$. Here, $A$ is the measured dE/dx, $B$ is the expected dE/dx and $\sigma_{TPC}$ is the dE/dx resolution as a function of particle track length ($L$) in the TPC. For pions, the $n\sigma_\pi$ distribution is basically close to a Gaussian with zero mean and unity sigma. The $m^2$ is calculated by $m^2 = p^2 \cdot ((t_{TOF} \cdot c/l)^2 - 1)$, where $p$ is the momentum, $t_{TOF}$ is the time of flight, $c$ is the speed of light in the vacuum, and $l$ is the flight path length of the particle. The pion, kaon and (anti-) proton peak can be clearly seen in Figure 2. The black line in Figure 2(b) is the projection of Figure 2(a) in the $m^2$ direction, in the transverse momentum bin of 3–4 GeV/c. At this $p_T$ range, the pion and kaon bands are merged together, and cannot be clearly separated from the (anti-) proton band. However, if we require $n\sigma_\pi > 0$ and then plot the $m^2$ distribution again (red line in Figure 2(b)), we find that the kaon and (anti-) proton bands are greatly suppressed and a clean pion signal is observed. This is due to the large difference between $n\sigma_\pi$ and $n\sigma_K$ (~ 1.7), or $n\sigma_{p(\bar{p})}$ (~ 2.2) in this $p_T$ range. Similarly, if we require $n\sigma_{p(\bar{p})} < 0$, the pion and kaon bands are suppressed significantly with respect to the (anti-) proton band, which help us get a cleaner (anti-) proton signal. Therefore, the combination of the TPC and TOF can extend the particle identification to higher momentum.

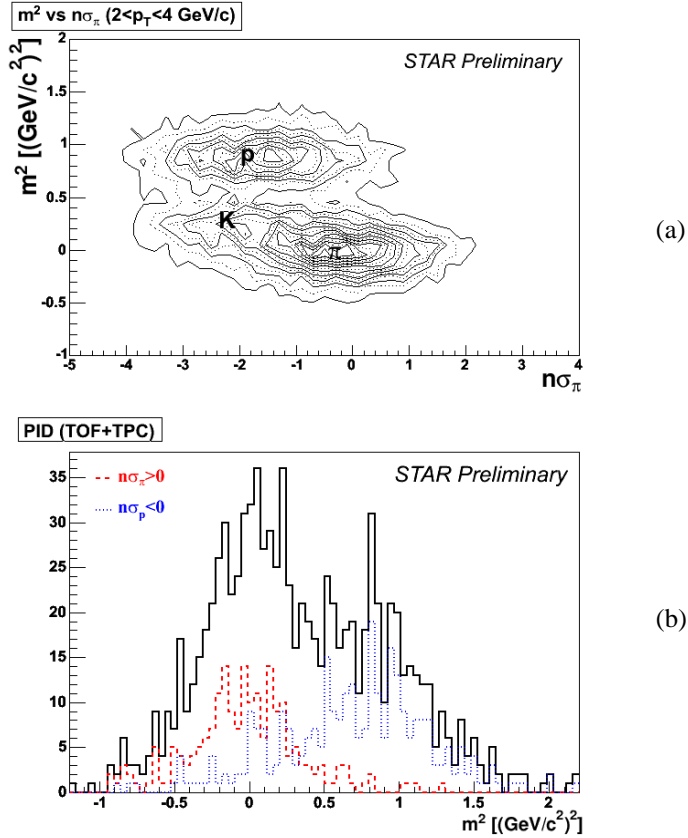

Figure 2: (a) Particle distribution as a function of $n\sigma_\pi$ and $m^2$. The pion, kaon and (anti-) proton peaks can be clearly seen. (b) $m^2$ distribution without dE/dx cut (black solid line), with $n\sigma_\pi > 0$ cut (red dash line), and $n\sigma_{p(\bar{p})} < 0$ cut (blue dotted line). The dE/dx cuts help getting a cleaner pion and (anti-) proton signals.

Before obtaining the yields of pions and (anti-) protons, the characteristics of $n\sigma_\pi$ or $n\sigma_{p(\bar{p})}$ must be determined. For example, if cutting on the $m^2$ distribution (as shown in Figure 3(a)) we can get a clean pion sample at $p_T < 2.5$ GeV/c. Figure 3(b) depicts the $n\sigma_\pi$ distribution of this sample fit to a Gaussian function. The mean and sigma of the fits as a function of $p_T$ are listed in Table 1. Since the values of the mean and the sigma are almost constant, one can extend this method to higher $p_T$ bins. The same procedure is also applied to get the parameters of $n\sigma_{p(\bar{p})}$. The final parameters of both $n\sigma_\pi$ and $n\sigma_{p(\bar{p})}$, for minimum bias events and different centrality bins in Au+Au collisions are summarized in Table 2. The centrality bins correspond to 0-10%, 10-20%, 20-40%, and 40-80%, with the size of the classes mainly driven by the limited statistics. The centrality is determined according to the event charged particle multiplicity measured in the TPC.

With the parameters of $n\sigma_\pi$ and $n\sigma_{p(\bar{p})}$ being determined, we can now calculate the yields of pions and (anti-) protons:

$$yield_\pi = A_\pi \cdot yield_\pi(n\sigma_\pi > 0) \quad \text{and}$$
$$yield_{p(\bar{p})} = A_{p(\bar{p})} \cdot yield_{p(\bar{p})}(n\sigma_{p(\bar{p})} < 0),$$

where the values of $A_\pi$ and $A_{p(\bar{p})}$ are listed in Table 1.

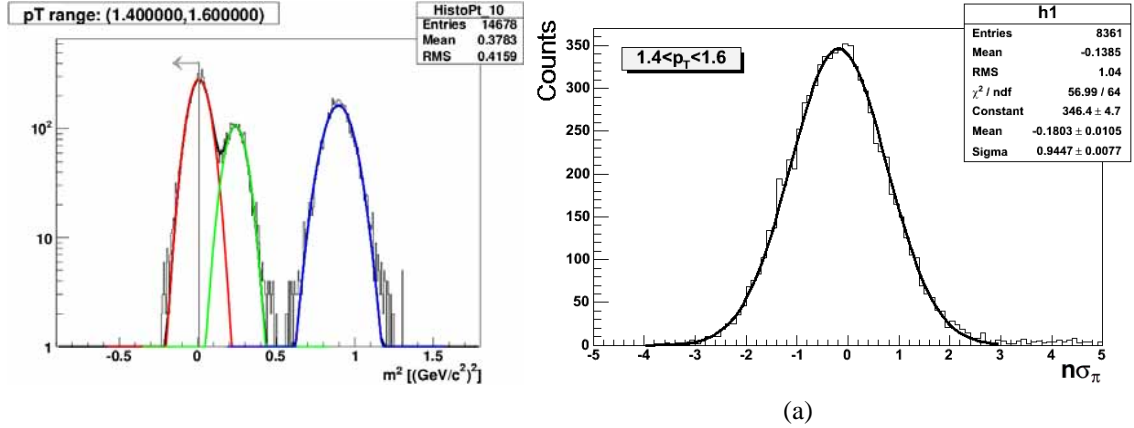

(a)

Figure 3: (a) $m^2$ cut at the pion mass in order to obtain a clean pion sample. (b) $n\sigma_\pi$ distribution using the pion sample resulting from the cut above. The solid lines are fits to a Gaussian function.

| $p_T$ range (GeV/c) | $n\sigma_\pi$ Gaussian parameters | |
|---|---|---|
| | Mean | Sigma |
| 0.9 – 1.0 | -0.194 | 0.981 |
| 1.0 – 1.2 | -0.200 | 0.970 |
| 1.2 – 1.4 | -0.185 | 0.943 |
| 1.4 – 1.6 | -0.186 | 0.949 |
| 1.6 – 1.8 | -0.209 | 0.915 |
| 1.8 – 2.0 | -0.256 | 0.916 |
| 2.0 – 2.5 | -0.194 | 0.962 |

Table 1: $n\sigma_\pi$ Gaussian fit parameters for different $p_T$ ranges.

| Centrality bins | Averaged $n\sigma_\pi$ Gaussian parameters | | $A_\pi$ | Averaged $n\sigma_{p(\bar{p})}$ Gaussian parameters | | $A_{p(\bar{p})}$ |
|---|---|---|---|---|---|---|
| | Mean | Sigma | | Mean | Sigma | |
| 0 - 10% | -0.36 | 0.97 | 2.76 | -0.33 | 0.99 | 1.59 |
| 10 – 20% | -0.23 | 0.94 | 2.48 | -0.25 | 0.96 | 1.66 |
| 20 – 40% | -0.11 | 0.92 | 2.20 | -0.09 | 0.98 | 1.87 |
| 40 – 80% | -0.07 | 0.93 | 2.13 | -0.07 | 0.95 | 1.89 |
| minbias | -0.2 | 0.95 | 2.40 | -0.2 | 1.0 | 1.73 |

Table 2: Averaged $n\sigma_\pi$ and $n\sigma_{p(\bar{p})}$ Gaussian parameters for minimum bias Au+Au collisions and 4 different centrality bins. The yield factors $A_\pi$ and $A_{p(\bar{p})}$ are also listed.

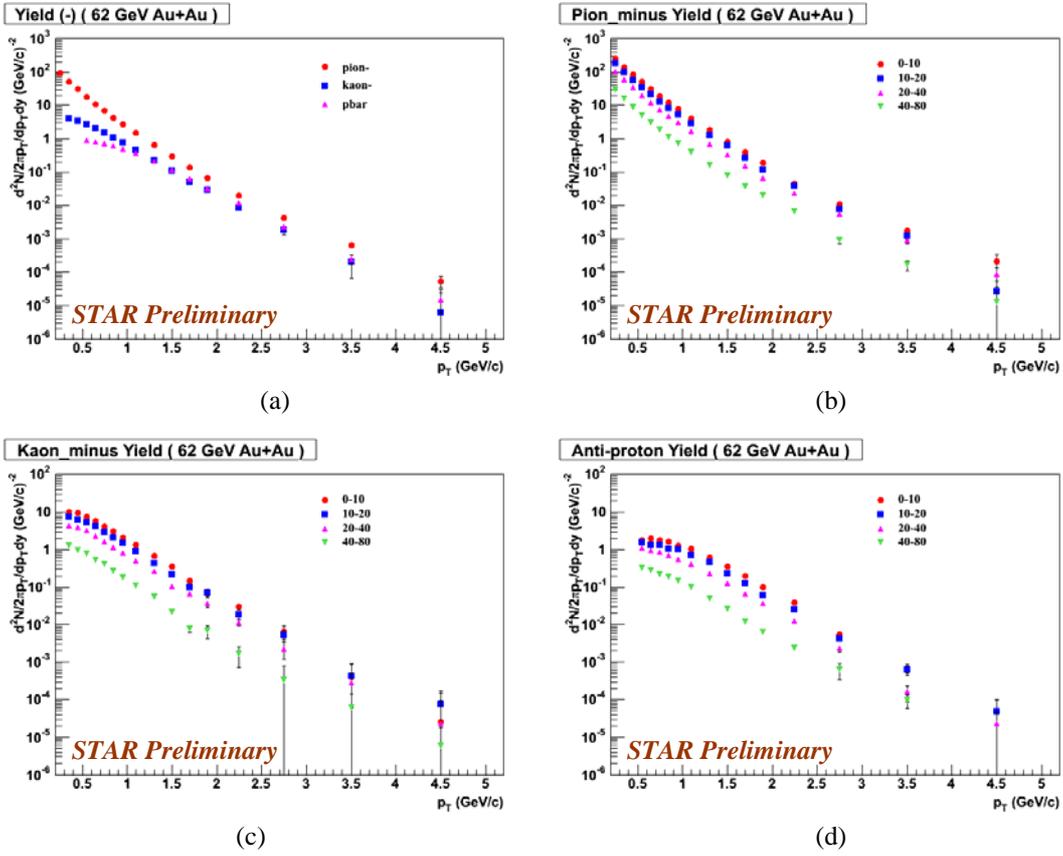

(a)  (b)  (c)  (d)

Figure 4: Invariant yields of $\pi^-$, $K^-$ and $\bar{p}$ as a function of the transverse momentum for minimum bias and different centralities in 62.4 GeV Au+Au collisions. The errors are statistical only.

## 4. Pion, Kaon and (Anti-) Proton Spectra

Using the method discussed above, the invariant yields of pions, kaons and (anti-) protons are calculated for 62.4 GeV Au+Au collisions. At the low $p_T$ region, the yields are obtained from a 3-Gaussian function fit to the $m^2$ spectra. While at larger $p_T$, the yields of pions and (anti-) protons are calculated applying the $n\sigma_\pi$ and $n\sigma_{p(\bar{p})}$ cuts, respectively. The kaon yield is obtained by subtracting the pion and (anti-) proton yields from the total hadron yield. The results are shown in Figure 4(a-d). The errors shown in these figures are statistical only. Only the yields of negative charged particles are presented here. The yields were corrected for detector acceptance and efficiency determined from a detailed simulation of the experimental apparatus. No background correction was applied to the spectra, and we estimate that the contribution is ~12% at low $p_T$ and ~5% at high $p_T$ for the pion yield, and ~20% for the (anti-) proton yield [18].

As can be seen in Figure 4, the pions and (anti-) protons are identified up to $p_T \sim$ 5 GeV/c, and the kaons are identified at least up to $p_T \sim$ 3 GeV/c.

## 5. Discussions and Conclusions

In Figure 5(a), the anti-particle to particle ratios ($\pi^-/\pi^+$, $K^-/K^+$ and $\bar{p}/p$) are shown as a function of $p_T$. The ratios basically do not change with $p_T$. A constant fit gives the overall anti-particle to particle ratio of $\pi^-/\pi^+=1.02\pm0.01$, $K^-/K^+=0.84\pm0.01$, and $\bar{p}/p=0.46\pm0.01$. The errors are statistical only. The $\bar{p}/p$ ratio reveals a significant net baryon excess in 62.4 GeV Au+Au collisions. The change of $p/\pi^+$ and $\bar{p}/\pi^-$ ratios as a function of $p_T$ is shown in Figure 5(b). A peak can be seen at $p_T \sim$ 2-2.5 GeV/c for both ratios. At higher $p_T$, the $p/\pi^+$ and $\bar{p}/\pi^-$ ratios seem to decrease. This is rather interesting and qualitatively agrees with the prediction of the coalescence/recombination model [8][9].

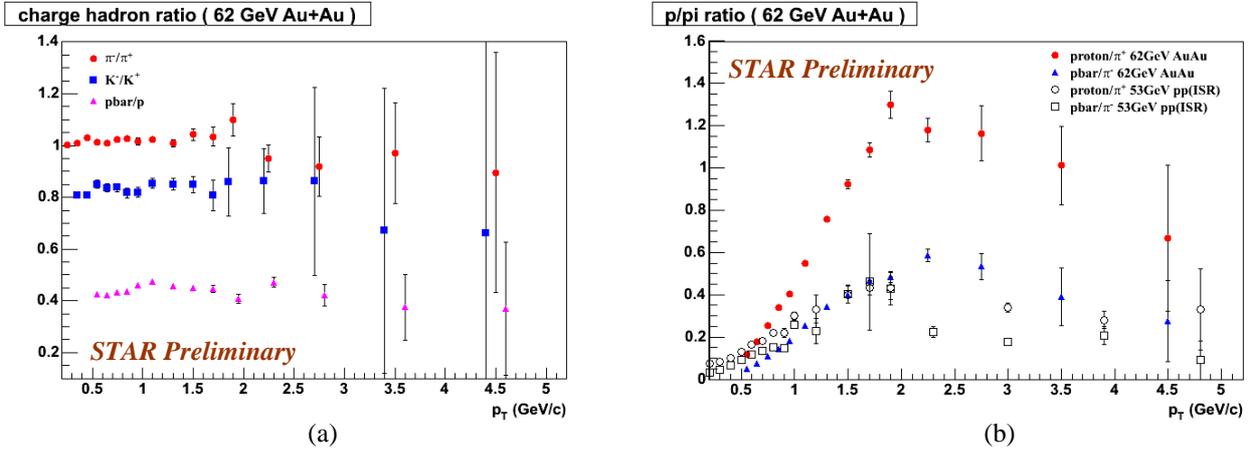

Figure 5: (a) $\pi^-/\pi^+$, $K^-/K^+$, and $\bar{p}/p$ ratios (b) $p/\pi^+$, $\bar{p}/\pi^-$ ratios as a function of $p_T$ in 62.4 GeV Au+Au collisions. Errors shown are statistical only. The $p/\pi^+$ and $\bar{p}/\pi^-$ ratios in 53 GeV pp collisions are also shown in (b) for comparison.

One may note that at $p_T$ above 3GeV/c and the significant difference between $n\sigma_\pi$ and $n\sigma_K$, the method discussed in section 3 can be used without any information from TOF. The TOF is used only to prove the validity of this method. Therefore, we can identify pions to even higher $p_T$, as long as $n\sigma_\pi$ and $n\sigma_K$ can be well sepa-rated ($\sim 2\sigma$) as demonstrated in Figure 6(a). The pion spectrum at higher $p_T$ using this method is shown in Figure 6(b). The spectrum obtained using the TOFr is also shown for comparison. It is clear that the results from the TPC only and TOFr are consistent in the overlapping $p_T$ region, which again proves the feasibility of this method.

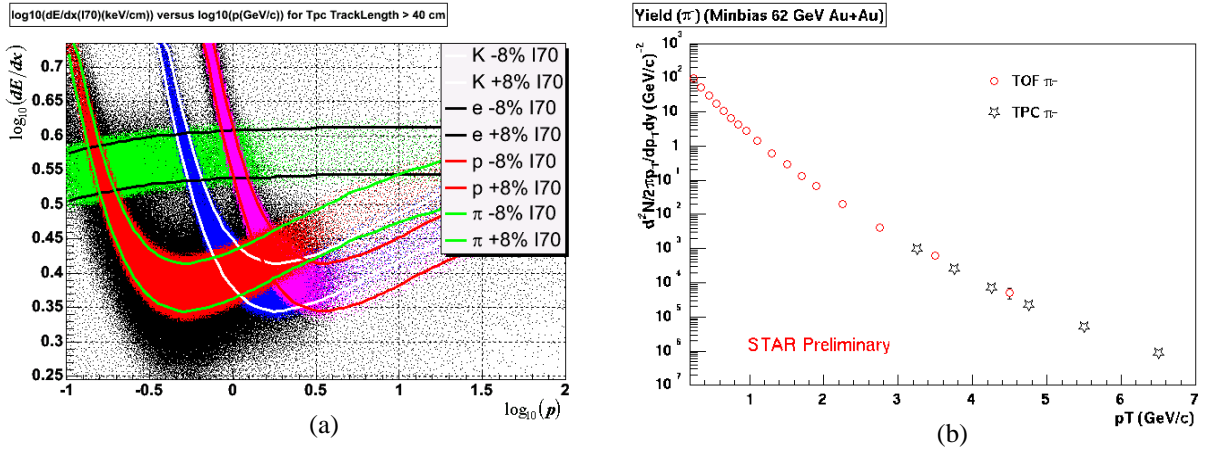

Figure 6: (a) Distribution of $\log_{10}(dE/dx)$ as a function of $\log_{10}(p)$ for electrons, pions, kaons and (anti-) protons. The units of dE/dx and momentum (p) are keV/cm and GeV/c, respectively. The color bands denote within $\pm 1\sigma$ the dE/dx resolution. (b) The invariant yield of $\pi^-$ from the TPC (at $p_T$ > 3 GeV/c only) and TOFr. Errors are statistical only.

In summary, we have developed a technique to extend the $p_T$ reach in the particle identification capability at STAR. By combining information from the TPC and TOF, we can measure pion, kaon and (anti-) proton spectra in the intermediate $p_T$ range. Preliminary spectra of pions up to $p_T \sim$ 7 GeV/c, (anti-) protons up to $p_T \sim$ 5 GeV/c, and kaons up to $p_T \sim$ 3 GeV/c were obtained in 62.4 GeV Au+Au collisions at RHIC. The anti-hadron to hadron ratios as the function of $p_T$ were presented. The overall $\bar{p}/p$ of $0.46 \pm 0.01$ reveals a significant net baryon excess in 62.4 GeV Au+Au collisions. The $p/\pi^+$ and $\bar{p}/\pi^-$ ratios have a maximum at $p_T \sim$ 2-2.5 GeV/c and then decrease at higher $p_T$.